\documentclass[a4paper,10pt]{article}

\usepackage{amssymb,amsmath,color,cite,graphicx,soul}
\title{On the propagation of Dirac fermions in graphene with the strain-induced inhomogeneous Fermi velocity}
\author{Alonso Contreras-Astorga$^1$, V\'i{}t Jakubsk\'y$^2$, Alfredo Raya$^3$\\\vspace{1mm}\\
$^1$\textit{CONACYT-Departamento de F\'isica, Cinvestav, A.P. 14-740, 07000 Ciudad de M\'exico, Mexico}\\
$^2$\textit{The Czech Academy of Science, Nuclear Physics Institute, Rez/Prague, Czech
		Republic}\\
$^3$\textit{ Instituto de F\'{\i}sica y Matem\'aticas, Universidad Michoacana de San Nicol\'as de Hidalgo.}\\\textit{Edificio C-3, Ciudad Universitaria. Francisco J. M\'ijica s/n Col. Fel\'{\i}citas del R\'{\i}o.}\\ \textit{ C. P. 58040, Morelia, Michoac\'an, Mexico.}
\\\sl{\small{E-mails: alonso.contreras@conacyt.mx,  
		jakubsky@ujf.cas.cz, alfredo.raya@umich.mx}} }

\begin{document}
	\maketitle
	\begin{abstract}
	We consider systems described by the two-dimensional Dirac equation where the Fermi velocity is inhomogeneous as a consequence of mechanical deformations. We show that the mechanical deformations can lead to deflection and focusing of the wave packets. The analogy with known reflectionless quantum systems is pointed out. Furthermore, with the use of the qualitative spectral analysis, we discuss how inhomogeneous strains can be used to create waveguides for valley polarized transport of partially dispersionless wave packets. 
	\end{abstract}
\section{Introduction}
Dirac fermions in graphene cannot be controlled very well by electrostatic field as they can tunnel through electrostatic barriers~\cite{geim}. Strain engineering~\cite{pereira} (also called straintronics or origami electronics~\cite{tomanek}), attracts increasing attention as a viable option for the design of electronic devices via mechanical deformations~\cite{pereira}-\cite{wupeeters}. Indeed, strains or folds of graphene sheet can be used to create waveguides~\cite{pereira}, \cite{wupeeters}, \cite{villegas}, \cite{zhai}, \cite{wu}. They can also lead to collimation, focusing or valley polarization of electron beams ~\cite{pereira}, \cite{wupeeters}, \cite{flouris}, or to the Kondo effect \cite{zhaikondo}. 

In graphene, the mechanical deformation of the crystal is manifested by the appearance of gauge fields \cite{vozmedianogauge}, \cite{naumisterronesreview} in an effective Dirac Hamiltonian and, in this way, it affects the dynamics of Dirac fermions. 
The mechanical deformation is expressed by strain tensor $\mathbf{u}=\mathbf{u}(x,y)$. It is given in terms of the displacement vector $u=(u_1(x,y),u_2(x,y))$ and vertical displacement $h=h(x,y)$,
\begin{equation}\label{straintensor}
\mathbf{u}_{ij}=\frac{1}{2}\left(\partial_iu_j+\partial_ju_i+\partial_jh\partial_ih\right).
\end{equation}
The Hamiltonian of the quasi-particle with the momentum in the vicinity of the Dirac point $\mathbf{K}$ then reads\footnote{$\partial_1\equiv\partial_x$, $\partial_2\equiv\partial_y$.} 
\begin{equation}\label{H}
H=-i\hbar\sigma_i\sqrt{\mathbf{v}_{ij}}\partial_j\sqrt{\mathbf{v}_{ij}}+v_0\hbar\sigma_iA_i+V,\quad i,j\in\{1,2\}.
\end{equation}
The tensor of Fermi velocity is defined as \cite{naumis}
\begin{equation}\label{vf1}
\mathbf{v}_{ij}=v_0\left(\mathbf{\eta}_{ij}+(1-\beta)\mathbf{u}_{ij}-\frac{1}{2}\partial_i h\partial_j h\right),
\end{equation}
whereas the vector potential and the electrostatic potential $V$ that emerge due to the mechanical deformation are
\begin{equation}\label{A}
A_1=\frac{\beta}{2a}(\mathbf{u}_{11}-\mathbf{u}_{22}),\quad A_2=\frac{\beta}{2a}(-2\mathbf{u}_{12}),
\end{equation}
\begin{equation}\label{V}
V=g(\mathbf{u}_{11}+\mathbf{u}_{22}).
\end{equation}
Here $\mathbf{\eta}=diag(1,1)$, $\beta$ is the electron Gr\"uneisen parameter and $a$ is the interatomic distance and $v_0$ the Fermi velocity in the strain-free crystal ($v_0\sim10^6\rm m/s$, $\beta\sim 2-3$ and $ a=1.46\AA$  for graphene). The bare value of the coupling constant $g$ has not been fixed definitely yet. In the literature, its value range between $0\,\rm eV$ to $\sim20\,\rm eV$, see e.g. \cite{vozmedianogauge}, \cite{zhai}. However, its magnitude seems to be considerably decreased due to screening, even up to the point that it renders the potential $V$ negligible \cite{sohier}. In this work, we set $\hbar=1$.

The strain in graphene, as well as in the other two dimensional materials, can be achieved by putting the material on the substrate that is micro-structured \cite{yingjie} or mechanically deformed \cite{roldan}, \cite{zhanggang}. The strain can also appear due to the mismatch of the lattices of the material and the substrate that gives rise to superlattices and associated Moire patterns~\cite{artaud}. It is worth mentioning that there are two-dimensional  systems where the Fermi velocity of Dirac fermions is intrinsically inhomogeneous~\cite{organic}, \cite{molecular}. Let us also mention models for corrugated graphene based on the hibridization of electron orbitals  \cite{pudlak} where the effect of the deformation is manifested by electrostatic potential.

The Hamiltonian (\ref{H}) resembles the energy operator in presence of electromagnetic field. However, it contains the inhomogeneous Fermi velocity that arises due to the shift of the Dirac points caused by the mechanical deformation \cite{naumis}, \cite{vozmendiano}, \cite{ramezani}, \cite{iorio}, \cite{volovik}. The formula (\ref{vf1}) for Fermi velocity is obtained when the tight-binding Hamiltonian is linearized around the shifted Dirac point \cite{naumis},\cite{volovik}.

In our work, we focus on two specific situations described by the Hamiltonian (\ref{H}) where the inhomogeneous strain tensor gives rise to the diagonal Fermi velocity with position dependent components. In both cases, we neglect the strain-induced electric potential (\ref{V}). We suppose that it vanishes either because of screening or it gets compensated by an external electric field.
In the next section, we consider the system with diagonal Fermi velocity whose upper component is $x$-dependent whereas its second non-vanishing component is constant. We show that the trajectory of the wave packets gets deflected by inhomogeneous strains. The trajectory can be obtained analytically with the use of a specific transformation that relates the considered system with the free particle model. It is worth noticing in this context that there were discussed systems in the literature where propagation of Dirac fermions manifested analogies with the optical systems. Let us mention \cite{wupeeters}, \cite{cao}, \cite{yesilyurt}, \cite{zhai2}, \cite{zhang} where it  was discussed the scattering of Dirac fermions on the barriers induced by  strain in combination with external fields. There, the quantum analog of Goos-H\"anschen effect was analyzed as well as possible valley polarization of the incoming electron beam. 

In the third section, we consider the system with diagonal strain tensor whose upper component depends on $x$, while its lower component is $y$-dependent. We focus on the analysis of the confinement of Dirac fermions within the wave guide formed by the strain. In literature, there have been already considered explicit models with piecewise constant strains \cite{pereira}, \cite{wupeeters}, \cite{villegas}, or smooth deformation profiles \cite{zhai}, \cite{wu}, where the spectrum of models was found numerically, see also \cite{wu} for experimental results. We apply another approach based on the qualitative spectral analysis of Dirac equation. It does not require the knowledge of solutions of the stationary equation and the explicit form of the strain-induced barrier is not essential. We show that the strained system is related to the strain-free model with external magnetic field. Existence of partially dispersionless wave packets in the wave guide is discussed. We use simple criteria  \cite{VJDK}, \cite{MFVJMT} to find the strain configurations that lead to appearance of these guided modes. The last section is left for discussion.

\section{Deflection of wave packets by mechanical deformations}
First, let us consider the strain tensor and the associated Fermi velocity in the following form,
\begin{equation}\label{v1}
\mathbf{u}=\left(\begin{array}{cc}\mathbf{u}_{11}(x)&0\\0&0\end{array}\right),\quad\mathbf{v}=\left(\begin{array}{cc}\mathbf{v}_{11}(x)&0\\0&v_0\end{array}\right).
\end{equation}
We suppose that $\mathbf{u}_{11}(x)$ is a bounded positive function, $\lim_{x\rightarrow\pm\infty}\mathbf{u}_{11}'(x)=0$, such that $\mathbf{v}_{11}(x)$ is also bounded and positive with a bounded derivative, $\lim_{x\rightarrow\pm\infty}\mathbf{v}_{11}'(x)=0$. 
The inhomogeneous strain and Fermi velocity (\ref{v1}) can be induced by a unidirectional strain and/or vertical displacements (\ref{vf1}).
The corresponding equation of motion for the two-dimensional Dirac fermion is 
\begin{equation}\label{eq1}
 i\partial_t\Psi(x,y,t)=H(x,y)\Psi(x,y,t)=	\left(-i\sqrt{\mathbf{v}_{11}(x)}\sigma_1\partial_x\sqrt{\mathbf{v}_{11}(x)}-iv_0\sigma_2\partial_y+\sigma_1\frac{v_0\beta}{2a}\mathbf{u}_{11}(x)\right)\Psi(x,y,t). 
\end{equation}
The Hamiltonian can be transformed to the free particle energy operator. Let us have the equation 
\begin{equation}
H_0(r,s)\Phi(r,s,t)=v_0(-i\sigma_1\partial_r-i\sigma_2\partial_s)\Phi(r,s,t)=i\partial_t\Phi(r,s,t).
\end{equation}	
We define transformation of coordinates
\begin{equation}\label{coordinates2}
r=\sin\gamma\, y+\cos\gamma\,\int_{0}^{x} \frac{v_0}{\mathbf{v}_{11}(q)}dq,\quad s=\cos\gamma\, y-\sin\gamma\,\int_{0}^x \frac{v_0}{\mathbf{v}_{11}(q)}dq,\quad \gamma\in (0,\pi/2)
\end{equation}
and the unitary operator
\begin{equation}
U(x)=\exp\left(i\sigma_3\frac{\gamma}{2}+\frac{i\,v_0\,\beta}{2a}\int^x_0 \frac{\mathbf{u}_{11}(q)}{\mathbf{v}_{11}(q)}dq\right).
\end{equation}
Then we have
\begin{equation}
H(x,y)=\mathbf{v}_{11}(x)^{-1/2}U^{-1}(x)H_0(r(x,y),s(x,y))U(x)\mathbf{v}_{11}(x)^{1/2}.
\end{equation}	
The eigenstates transform as
\begin{equation}
\Psi(x,y,t)=\left(\frac{\mathbf{v}_{11}(x)}{v_0}\right)^{-1/2}U^{-1}(x)\Phi(r(x,y),s(x,y),t),\quad H(x,y)\Psi (x,y,t)=i\partial_t\Psi(x,y,t).
\end{equation}
It can be verified by direct calculation that the transformation preserves the norm of the eigenstates,
\begin{equation}\label{norm}
\int_{\mathbb{R}^2}|\Psi(x,y,t)|^2dxdy=\int_{\mathbb{R}^2}\left|\left(\frac{\mathbf{v}_{11}(x)}{v_0}\right)^{-1/2}\Phi(r(x,y),s(x,y),t)\right|^2dxdy=\int_{\mathbb{R}^2}|\Phi(r,s,t)|^2drds.
\end{equation}
It is worth noticing that the transformation discussed in \cite{naumis} coincides with the one discussed here for $\gamma=0$.

Let us use the stationary states $\Phi_{k_r,k_s}$ of $H_0$,
\begin{equation}
H_0\Phi_{k_r,k_s}=\sqrt{k_r^2+k_s^2}\Phi_{k_r,k_s},\quad \Phi_{k_r,k_s}(r,s)=e^{i(k_r r+k_s s)}\left(1, \frac{k_r+ik_s}{\sqrt{k_r^2+k_s^2}}\right)^T,
\end{equation}
to compose a Gaussian wave packet 
\begin{equation}
\Phi(r,s,t)=\int_{\mathbb{R}^2} dk_r dk_s A(k_r,k_s)(\Phi_{k_r,k_s}(r,s)+\Phi_{k_r,-k_s}(r,s))e^{-i\sqrt{k_r^2+k_s^2}t}.
\end{equation}
We fix the coefficient function as $A(k_r,k_s)=e^{-\frac{(k_r-K_r)^2}{\sigma_r}-\frac{k_s^2}{\sigma_s}}$. Then the wave packet propagates along the $r$ axis ($s=0$) and disperses symmetrically with respect to the the $s$ axis. By definition, the wave packet satisfies $i\partial_t\Phi(r,s,t)=H_0(r,s)\Phi(r,s,t)
$. We can transform $\Phi(r,s,t)$ into the wave packet $\Psi(x,y,t)$ that would satisfy $(-i\partial_t+H(x,y))\Psi(x,y,t)=0$,
\begin{equation}\label{psievolves}
\Psi(x,y,t)=\left(\frac{\mathbf{v}_{11}(x)}{v_0}\right)^{-1/2}U^{-1}(x)\Phi(r(x,y),s(x,y),t).
\end{equation}
This wave packet does not longer move along a straight line, but rather follows the curve $s(x,y)=0$ corresponding to 
\begin{equation}y(x)=\tan \gamma\int_{0}^{x} \frac{v_0}{\mathbf{v}_{11}(q)}dq. \label{y}\end{equation}
It corresponds to the trajectory of a point particle moving with the velocity $v=(\mathbf{v}_{11}(x),v_0)$.
Let us discuss the trajectory of the wave packet in dependence on several configurations of the strain $\mathbf{u}_{11}$ in more detail.

\subsubsection*{\bf Asymptotically constant Fermi velocity}
\begin{center}
	\begin{figure}
		\begin{center}
			\includegraphics[scale=.55]{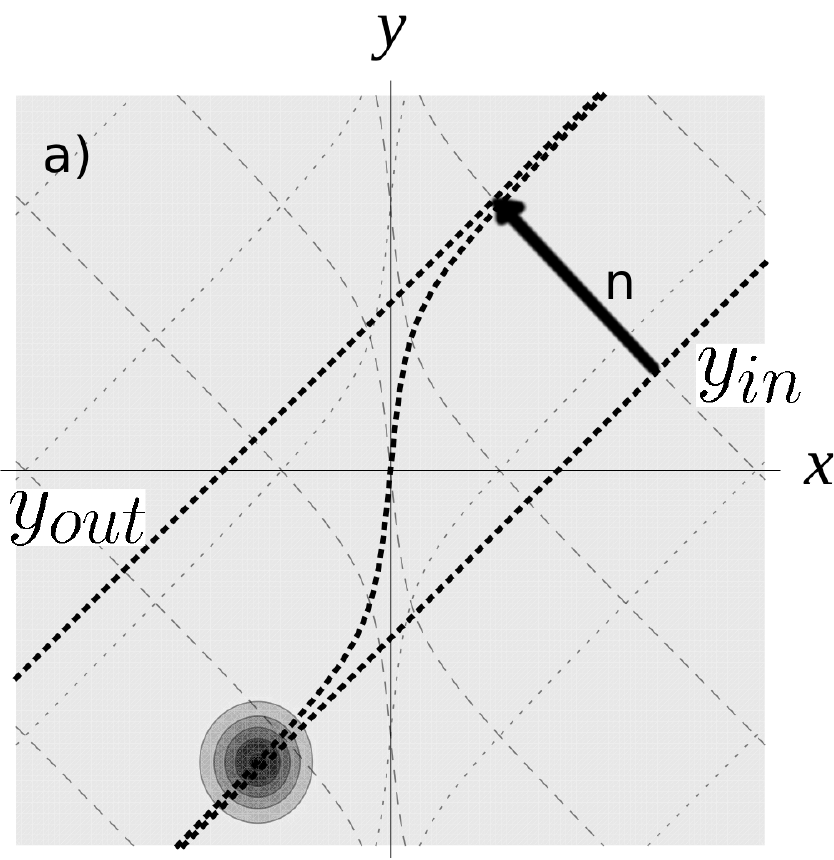}\hspace{2mm}
			\includegraphics[scale=.55]{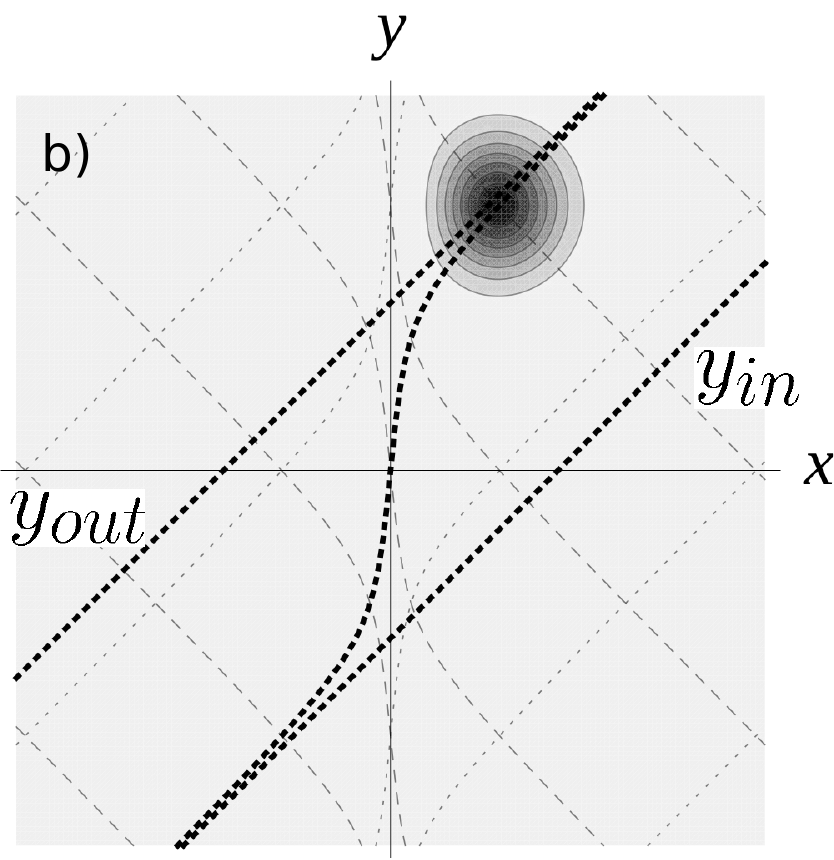}\hspace{2mm}		
		\end{center}
		\caption{Deflection of the wave packet by localized deformation. Probability density of a  wave packet at two subsequent instants of time a) $t_1$ and b) $t_2$ ($t_1<t_2$). The trajectory of the wave packet is determined by the two parallel lines $y_{in}$ and $y_{out}$, see (\ref{yinyout}), that represent incoming and outgoing trajectories. The deflection is quantified by the distance of $y_{in}$ and $y_{out}$ measured by the vector $n$, see (\ref{n}). The dotted curves correspond to constant values of $s$ whereas the dashed curves are given by fixed values of $r$. Thick black dashed curve is the trajectory of the wave packet. In the illustration, we combined the generic dispersing Gaussian wave packet with the  probability density  $|\Phi(r,s,t)|=\frac{1}{1+\sigma^2t^2}\exp\left(-\sigma\frac{(r-r_0-K_rt)^2+(s-s_0-K_st^2)}{4(1+\sigma^2t^2)}\right)$ with the formula (\ref{psievolves}).  }\label{bla0}
	\end{figure}
\end{center}

First, let us suppose that the Fermi velocity is asymptotically equal to $v_0$ and the strain induces only a localized fluctuation of $\mathbf{v}_{11}$,
\begin{equation}\label{vfluctuation}
\mathbf{v}_{11}=v_0(1+\Delta v),\quad \lim_{x\rightarrow\pm\infty}\Delta v=0.
\end{equation}
Deformation of this kind can be produced by folds of graphene sheet, see e.g. \cite{zhai}.
We suppose that the fluctuation $\Delta v$ is small in the following sense,
\begin{equation}
\int_{0}^{\infty}  \frac{\Delta v(q)}{1+\Delta v(q)}dx=X_+,\quad -\int_{-\infty}^0  \frac{\Delta v(q)}{1+\Delta v(q)}dx=X_-,\quad |X_{\pm}|<\infty. 
\end{equation}
Then the trajectory defined in (\ref{y}) tends asymptotically to two straight lines
\begin{equation}\label{yinyout}
y(x)\rightarrow\begin{cases}y_{in}=\tan\gamma\, (x-X_-),\quad x\rightarrow-\infty\\
y_{out}=\tan\gamma\, (x-X_+),\quad x\rightarrow\infty.\end{cases}
\end{equation}
The two asymptotic trajectories $y_{in}$ and $y_{out}$ are parallel but mutually shifted.  Therefore, the wave packet traveling along the trajectory (\ref{y}) gets deflected by the mechanical deformation that gives rise to the inhomogeneous Fermi velocity (\ref{vfluctuation}). It is straightforward to compute explicitly the length of the normal vector $n$ connecting the two lines, 
\begin{equation}\label{n}
|n|=\sin\gamma|X_+-X_-|.
\end{equation} 
see Fig.\ref{bla0} for illustration.

Now, we shall consider the situation where $\mathbf{v}_{11}(x)$ tends asymptotically to two, possibly different, constant values. It can be induced by the unidirectional strain which vanishes for $x\rightarrow-\infty$ but it converges to a constant positive value for $x\rightarrow\infty$. In this case, the displacement of the atoms in the crystal have linear-like growth for large values of $x$. We suppose that such a deformation can be achieved by the corresponding strain of the substrate on which the graphene sheet is positioned. 

\begin{center}
	\begin{figure}
		\begin{center}
			\includegraphics[scale=.52]{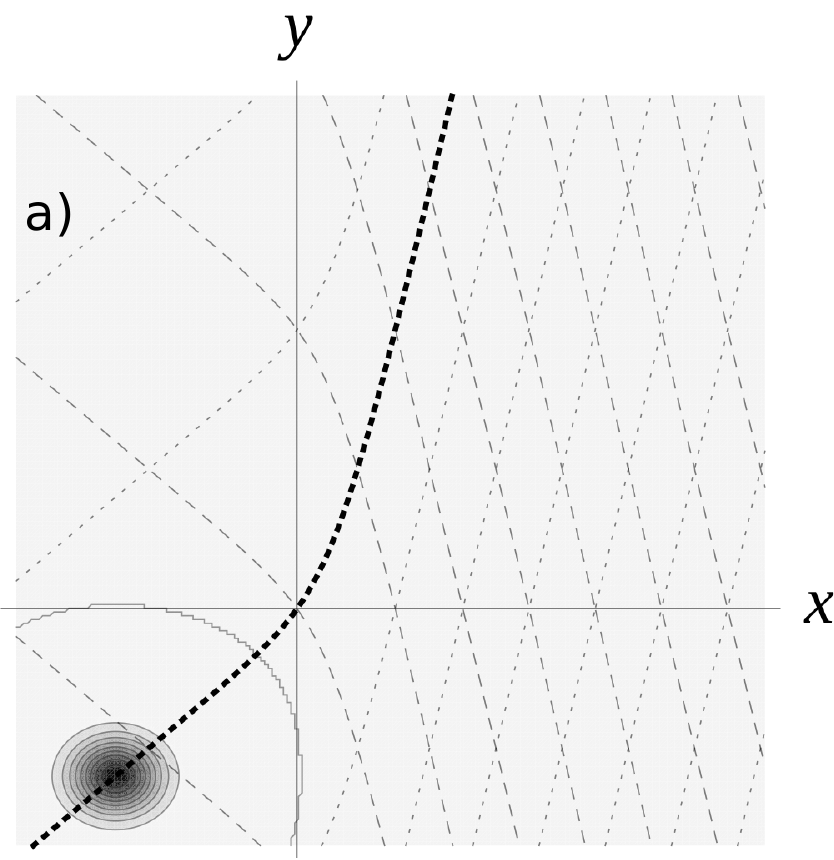}\hspace{2mm}
			\includegraphics[scale=.52]{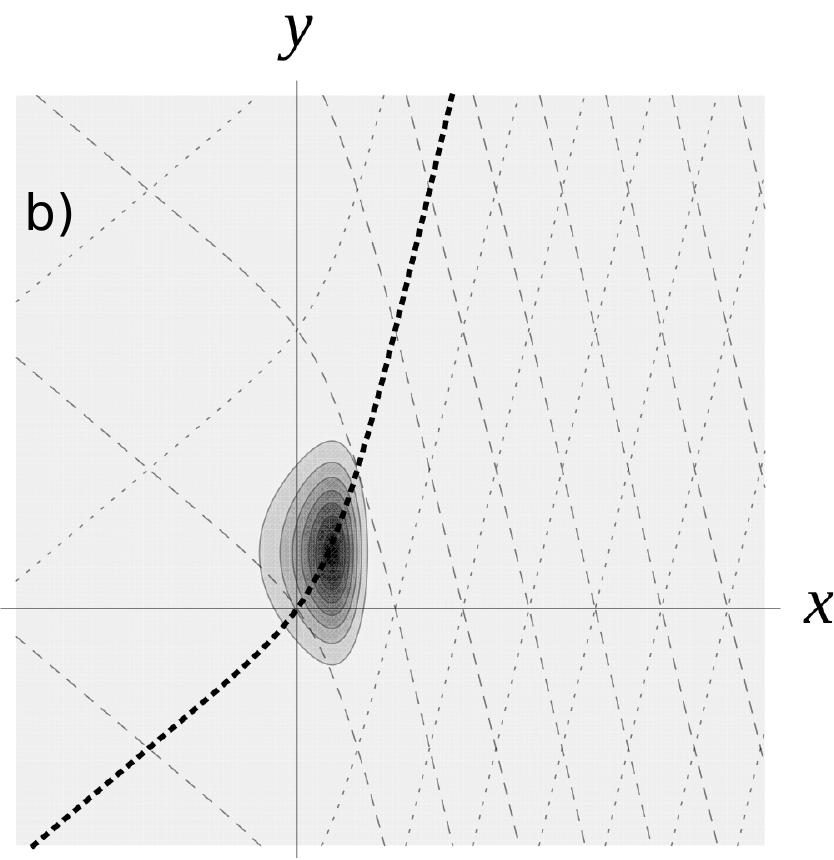}\hspace{2mm}		
			\includegraphics[scale=.52]{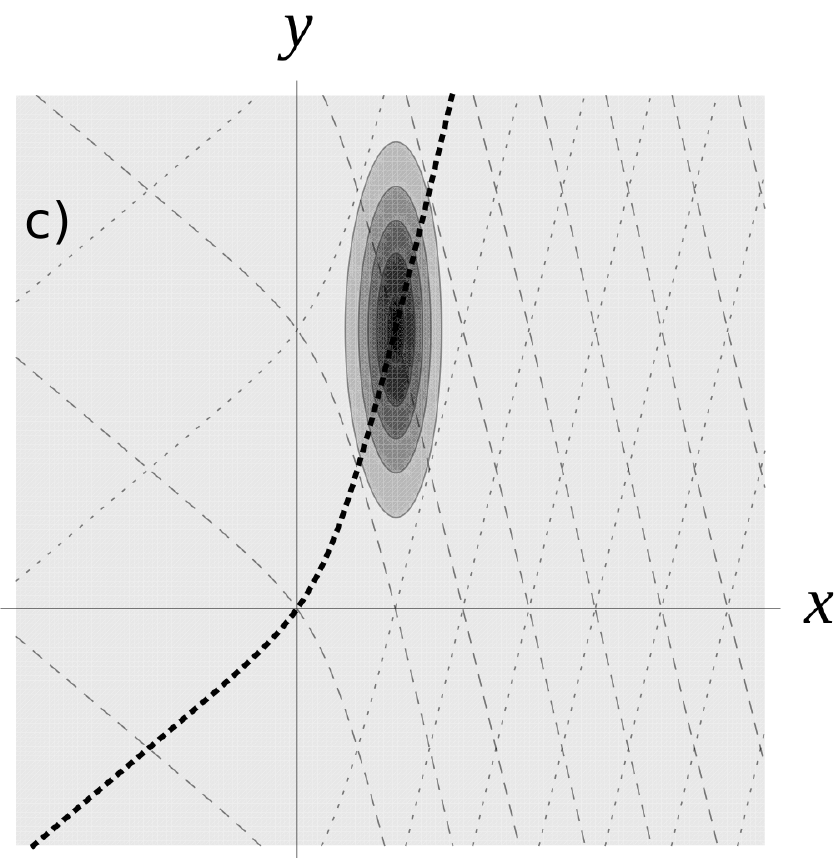}\\
			\includegraphics[scale=.52]{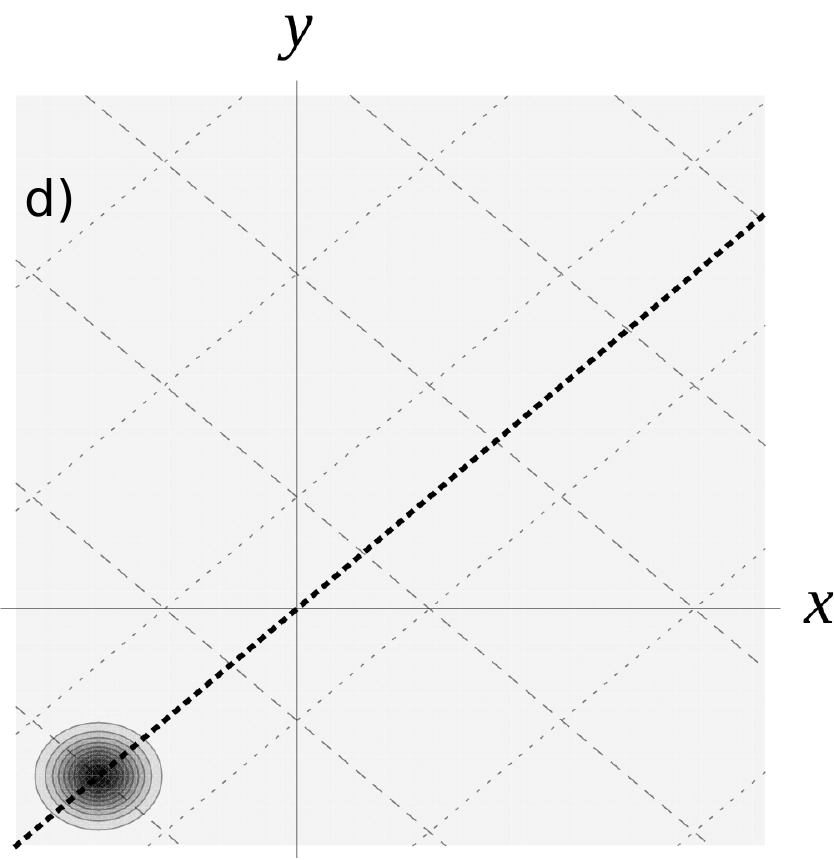}\hspace{2mm}
			\includegraphics[scale=.52]{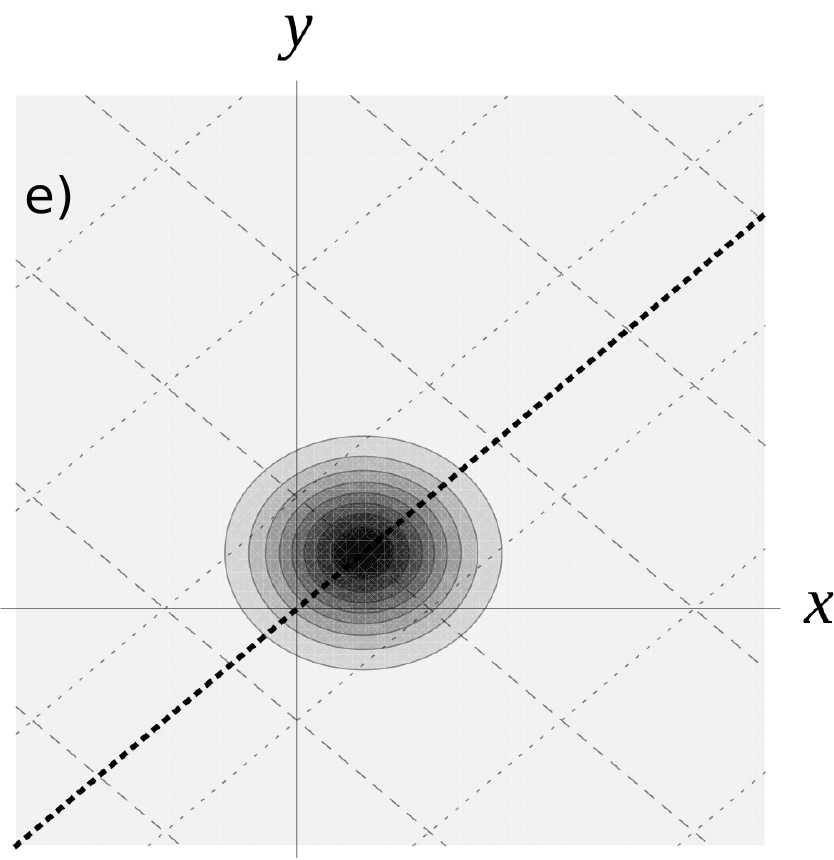}\hspace{2mm}
			\includegraphics[scale=.52]{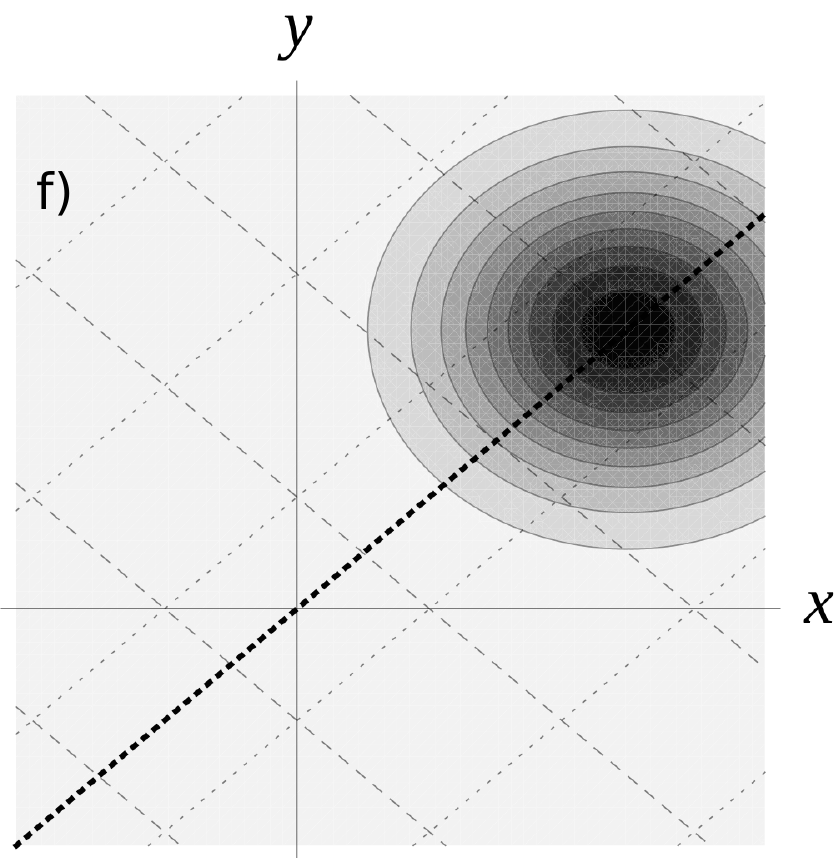}
		\end{center}
		\caption{Focusing of the probability density of a wave packet propagating through deformed crystal at three subsequent instants of time, a) $t=t_1$, b) $t=t_2$, c) $t=t_3$. For the illustration, we used $\mathbf{v}_{11}=v_0(1-\mu(1+\tanh\nu x))$. For comparison, the figures d)-f) in the lower row  illustrates the evolution of the same wave packet in undeformed crystal at the same times. The thick dotted line is the trajectory along which the wave packet propagates. The dashed curves are defined by constant values of $r$ whereas the dotted curves correspond to constant values of $s$, see (\ref{rs}). The wave packet was obtained by substitution of $\Phi(r,s,t)$ in (\ref{psievolves}) by a generic dispersing wave packet satisfying $|\Phi(r,s,t)|=\frac{1}{1+\sigma^2t^2}\exp\left(-\sigma\frac{(r-r_0-K_rt)^2+(s-s_0-K_st^2)}{4(1+\sigma^2t^2)}\right)$. }\label{bla}
	\end{figure}
\end{center}

Let us suppose that the deformation should be such that the induced Fermi velocity satisfies
\begin{equation}
\frac{v_0}{\mathbf{v}_{11}}=\begin{cases}\omega_- +O(x^{-\alpha}),& x\rightarrow-\infty,\\
\omega_+ +O(x^{-\alpha}),& x\rightarrow\infty,\quad \alpha>1.\end{cases}
\end{equation}
For these values of $\alpha$, the $O(x^{-\alpha})$ term gives convergent contribution to the trajectory (\ref{y}) of the wave packet so that $y(x)$ can be written as
\begin{equation}
y=\begin{cases}
\omega_-\tan\gamma\, x+O(1),& x\rightarrow-\infty,\\
\omega_+\tan\gamma\, x+O(1),& x\rightarrow\infty. \end{cases}
\end{equation}
This type of deformation is illustrated in Fig. \ref{bla} where we fixed $\mathbf{v}_{11}=v_0(1-\mu(1+\tanh\nu x))$ for constant $\mu$ and $\nu$.

\subsubsection*{\bf Periodic fluctuation of Fermi velocity}
Let us focus on the situation where the Fermi velocity is constant for all $x<0$ and periodic for $x\geq 0$,
\begin{equation}
\mathbf{v}_{11}=\begin{cases}v_0,\quad x<0,\\ 
							\mathbf{v}_{11}(x+L)=\mathbf{v}_{11}(x),\quad x\geq 0.
				\end{cases} 
\end{equation}
We suppose that the corresponding periodic strain can be achieved by the flexural modes \cite{naumis} or by interaction with the substrate that gives rise to the Moire patterns \cite{naumisterronesreview}.
Then the trajectory (\ref{y}) can be written as
\begin{equation}\label{yperiodic}
y(x)=	\begin{cases} \tan\gamma\, x,\quad x<0,\\
\tan\gamma\, \left(\left[\frac{x}{L}\right]A+\int_{\left[\frac{x}{L}\right]L}^x\frac{v_0}{\mathbf{v}_{11}(q)}dq\right),\quad A=\int_{0}^L\frac{v_0}{\mathbf{v}_{11}(q)}dq,\quad x\geq 0,
		\end{cases} 
\end{equation}
where $[x]$ denotes integer part of $x$.
In general, $y(x)$ is no longer linear for $x\geq0$. However, it can be confined between two parallel lines. Utilizing $x-1\leq \left[x\right]\leq x$ together with positivity of $\mathbf{v}_{11}(x)$, we get
\begin{equation}
\tan\omega\, x-A\tan\gamma\leq y(x)\leq \tan\omega\, x+A \tan \gamma, 
\end{equation}
where 
\begin{equation}\label{omega}
\tan\omega=\frac{A}{L}\tan\gamma=\frac{1}{L}\tan\gamma \int_{0}^L\frac{v_0}{\mathbf{v}_{11}(q)}dq.
\end{equation}
Hence, the periodic strain deflects the wave packet  by the angle 
$\omega-\gamma.$
The deflection angle $\omega$ is just function of the incidence angle $\gamma$, periodicity $L$ and the integral $A=\int_{0}^L\frac{v_0}{\mathbf{v}_{11}(q)}dq$ so that it is the same for a family of Fermi velocities that share these quantities. It is worth noticing that $\omega$ also depends on the material constants $\beta$ and $a$ that stay hidden in definition of $\mathbf{v}_{11}$. Contrary to the systems studied in \cite{cao}, \cite{zhang} where the Goos-H\"anchen-like effect for transmitted electrons on strain and potential barrier was analyzed, the electrons in our model pass through the strain-induced barrier without any reflection. There is also no difference in propagation of the wave packets formed in the $K$ and $K'$ valleys.

\section{Waveguides by inhomogeneous unidirectional strains }
In this section, we consider system where the strain tensor and the Fermi velocity acquire the following form
\begin{equation}\label{bidirectionalstrain}
\mathbf{u}=\left(\begin{array}{cc}\mathbf{u}_{11}(x)&0\\0&\mathbf{u}_{22}(y)\end{array}\right),\quad \mathbf{v}_{ij}=v_0\left(1+(1-\beta)\mathbf{u}_{ij}-\frac{1}{2}\partial_i h\partial_j h\right)\mathbf{\eta}_{ij}.
\end{equation} 
We suppose that $\mathbf{u}_{11}$, $\mathbf{u}_{22}$ and $h$ are bounded and continuous and they are such that $\mathbf{v}_{11}$ and $\mathbf{v}_{22}$ are strictly positive.
The stationary equation with the Hamiltonian (\ref{H}) reads as
\begin{eqnarray}
H(x,y)\Psi(x,y)&=&\left(-i\sigma_1\sqrt{\mathbf{v}_{11}(x)}\partial_x\sqrt{\mathbf{v}_{11}(x)}-i\sigma_2\sqrt{\mathbf{v}_{22}(y)}\partial_y\sqrt{\mathbf{v}_{22}(y)}+\sigma_1\frac{v_0\beta}{2a}(\mathbf{u}_{11}(x)-\mathbf{u}_{22}(y))\right)\Psi(x,y)\nonumber\\&=&E\Psi(x,y).\label{h2}
\end{eqnarray}
We make the gauge transformation to eliminate the $x$-dependent potential term and to simplify kinetic term, 
\begin{eqnarray}\label{gauge}
\tilde{H}(x,y)&=&G(x,y)H(x,y)G(x,y)^{-1}  \nonumber \\
	&=&-i\sigma_1\mathbf{v}_{11}(x)\partial_x-i\sigma_2\mathbf{v}_{22}(y)\partial_y-\sigma_1 \frac{v_0\beta}{2a}\mathbf{u}_{22}(y),\quad G(x,y)=\frac{\sqrt{\mathbf{v}_{11}\mathbf{v}_{22}}}{v_0}\,e^{i\frac{v_0\beta}{2a}\int \mathbf{u}_{11}/\mathbf{v}_{11}dx}.
\end{eqnarray}	
We change the coordinates
\begin{equation}\label{rs}
r=\int_0^x\frac{v_0}{\mathbf{v}_{11}(q)}dq,\quad s=\int_{0}^y\frac{v_0}{\mathbf{v}_{22}(q)}dq.
\end{equation}
Notice that they are are similar to (\ref{coordinates2}) for $\gamma=0$. Then the stationary equation gets the following form  
\begin{equation}\label{h(r,s)}
\tilde{H}(r,s)\psi(r,s)=v_0(-i\sigma_1\partial_r-i\sigma_2\partial_s+\sigma_1 F(s))\psi(r,s)=E\psi(r,s),\quad F(s)\equiv-\frac{\beta}{2a}\mathbf{u}_{22}(y(s)).
\end{equation}
We require $r$ and $s$ to be mappings from $\mathbb{R}$ \textit{onto} $\mathbb{R}$. They should be invertible, monotonic functions of $x$ and $y$ (there holds $r'(x)\neq 0$ and $s'(y)\neq 0$). The derivatives $r'(x)$ as well as $s'(y)$ should be bounded. For convenience, we fix 
\begin{equation}
\lim_{x\rightarrow \pm\infty}r(x)=\pm\infty,\quad \lim_{y\rightarrow \pm\infty}s(y)=\pm\infty.
\end{equation}
It is granted that there is also an inverse function $y=y(s)$ that satisfies $y(s(y))=y$.

It is more convenient to analyze the equation (\ref{h(r,s)}) instead of (\ref{h2}) due to its simpler form.
We can take advantage of the translational invariance of the system and focus on  subspaces with a conserved value of the momentum $-i\partial_r$. We make the partial Fourier transformation
\begin{equation}
(\mathcal{F}\psi)(k,s)=\frac{1}{\sqrt{2\pi}}\int_{\mathbb{R}}e^{-ik r}\psi(r,s)dr.
\end{equation}
The Hamitonian $\tilde{H}(r,s)$ can be rewritten as direct integral
\begin{equation}\label{tildeh(k,s)}
\tilde{H}(r,s)=\int^{\oplus}_{\mathbb{R}}\tilde{H}_k(s),\quad \tilde{H}_{k}(s)=v_0(\sigma_1k-i\sigma_2\partial_s+\sigma_1F(s)).
\end{equation}
The direct integral can be understood as a generalization of the partial wave decomposition for the case where the conserved quantum number is not discretized but acquires values from a real interval. As the potential term $F(s)$ is bounded and continuous, the Hamiltonian $H_{k}(s)$ is self-adjoint on the space of functions that are square integrable together with their first derivative.

We are interested in the configurations of mechanical strain where the Hamiltonian $\tilde{H}_{k}(s)$ possesses discrete energies. The reason is that the discrete energy levels give rise to discrete energy bands in the spectrum of $\tilde{H}(r,s)$ and $H(x,y)$ that can be associated with existence of (partially) dispersionless wave packets \cite{VJMT}. Indeed, let us suppose that we can find the solution of
\begin{equation}\label{eq2}
\tilde{H}_{k}(s)\psi_{k,n}(s)=E_n(k)\psi_{k,n}(s),\quad \forall k\in \mathcal{I}_n\subset\mathbb{R},
\end{equation} 
where $E_n(k)$ is a discrete energy of $\tilde{H}_{k}(s)$ labeled\footnote{We label just the positive energies as the spectrum is symmetric with respect to zero.} by $n$, i.e. $\psi_{k,n}(s)$ is square integrable together with its first derivative. The intervals $\mathcal{I}_n$ can be finite but also (semi-)infinite. 

\begin{center}
\begin{figure}
\begin{center}
	\includegraphics[scale=.4]{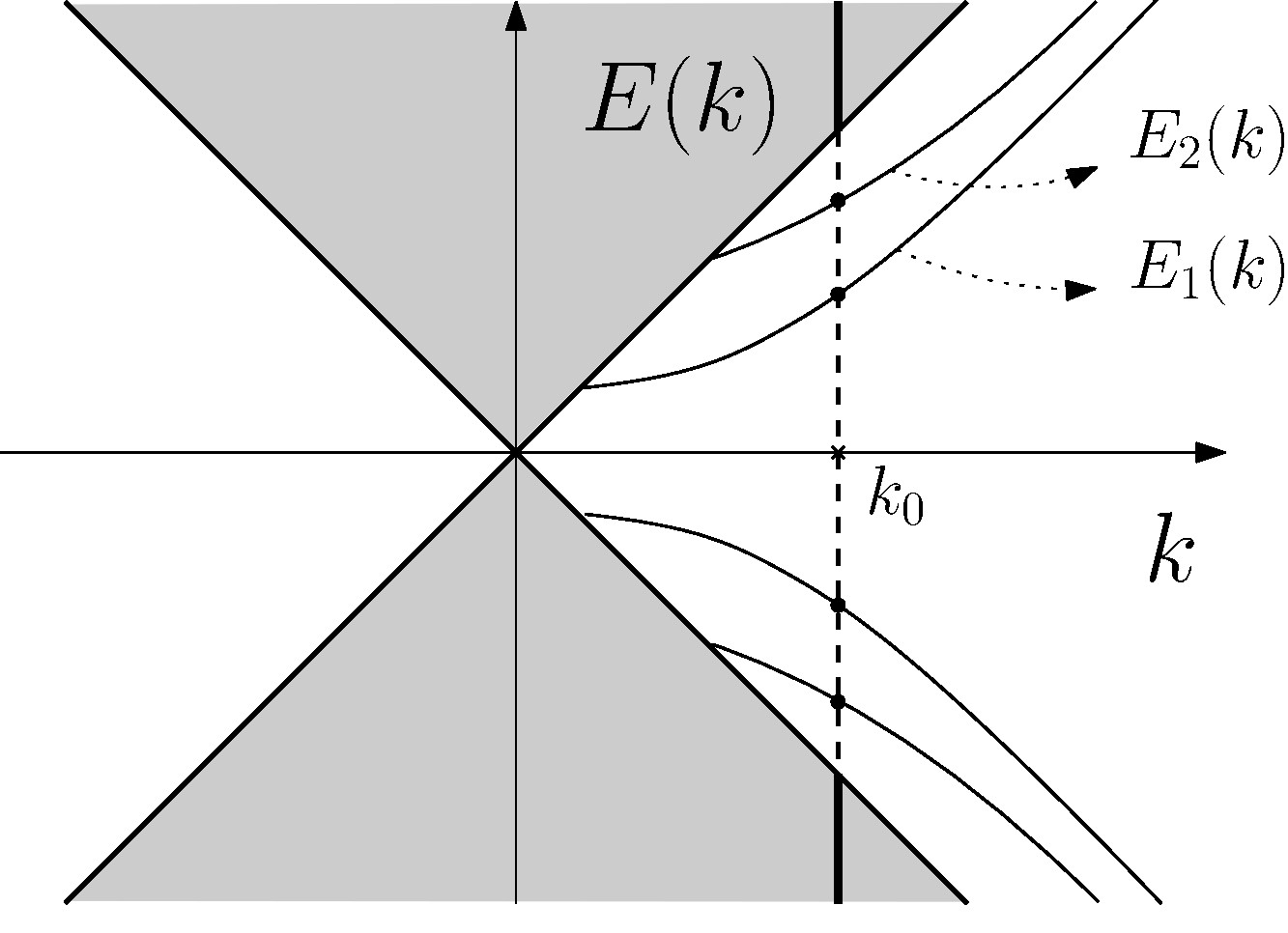}\hspace{8mm}
	\end{center}
	\caption{Spectrum of $H(x,y)$. For fixed $k_0$, we get energy spectrum of $H_k(s)$ that consists of discrete energies (black dots) and the bands of positive and negative energies. }\label{band}

	\end{figure}
\end{center}

We can get eigenstates of $H(x,y)$ from those of $\tilde{H}_k(s)$,
\begin{equation}\label{stacstate}
H(x,y)\Psi_{k,n}(x,y)=E_n(k)\Psi_{k,n}(x,y),\quad \Psi_{k,n}(x,y)=G(x,y)^{-1}e^{ikr(x)}\psi_{k,n}(s(y)).
\end{equation}
The discrete energies $E_n(k)$ form discrete energy bands in the spectrum of the two-dimensional Hamiltonian, see Fig.~\ref{band} for illustration.
We can use them to construct the following wave packets associated with the energy bands $E_n(k)$,
\begin{equation}\label{dispwp}
\Psi_n(x,y)=\frac{1}{\sqrt{2\pi}}\int_{\mathcal{I}_n}\rho_n(k)\Psi_{k,n}(x,y)dk.
\end{equation}
This wave packet is normalizable provided that $\rho(k)$ is normalizable (see Appendix for details),
\begin{eqnarray}
&&\int_{\mathbb{R}^2}\left|\Psi_n(x,y)\right|^2dxdy=\int_{\mathbb{R}}\left|\rho_n(k)\right|^2dk.
\end{eqnarray} 
The wave packet (\ref{dispwp}) has a remarkable property - it does not disperse along $y$ axis. Indeed, there holds (again, see Appendix for details)
\begin{eqnarray}
&&\int^{b}_ady\int_{\mathbb{R}}dx\left|e^{-itH(x,y)}\Psi_n(x,y)\right|^2=\int^{b}_ady \int_{\mathbb{R}}dx\left|\Psi_n(x,y)\right|^2,
\end{eqnarray}
where $a$ and $b$ are arbitrary real numbers, i.e. the probability density in the $y$ direction is conserved during the time evolution.

\begin{center}
\begin{figure}
\begin{center}
		\includegraphics[scale=.36]{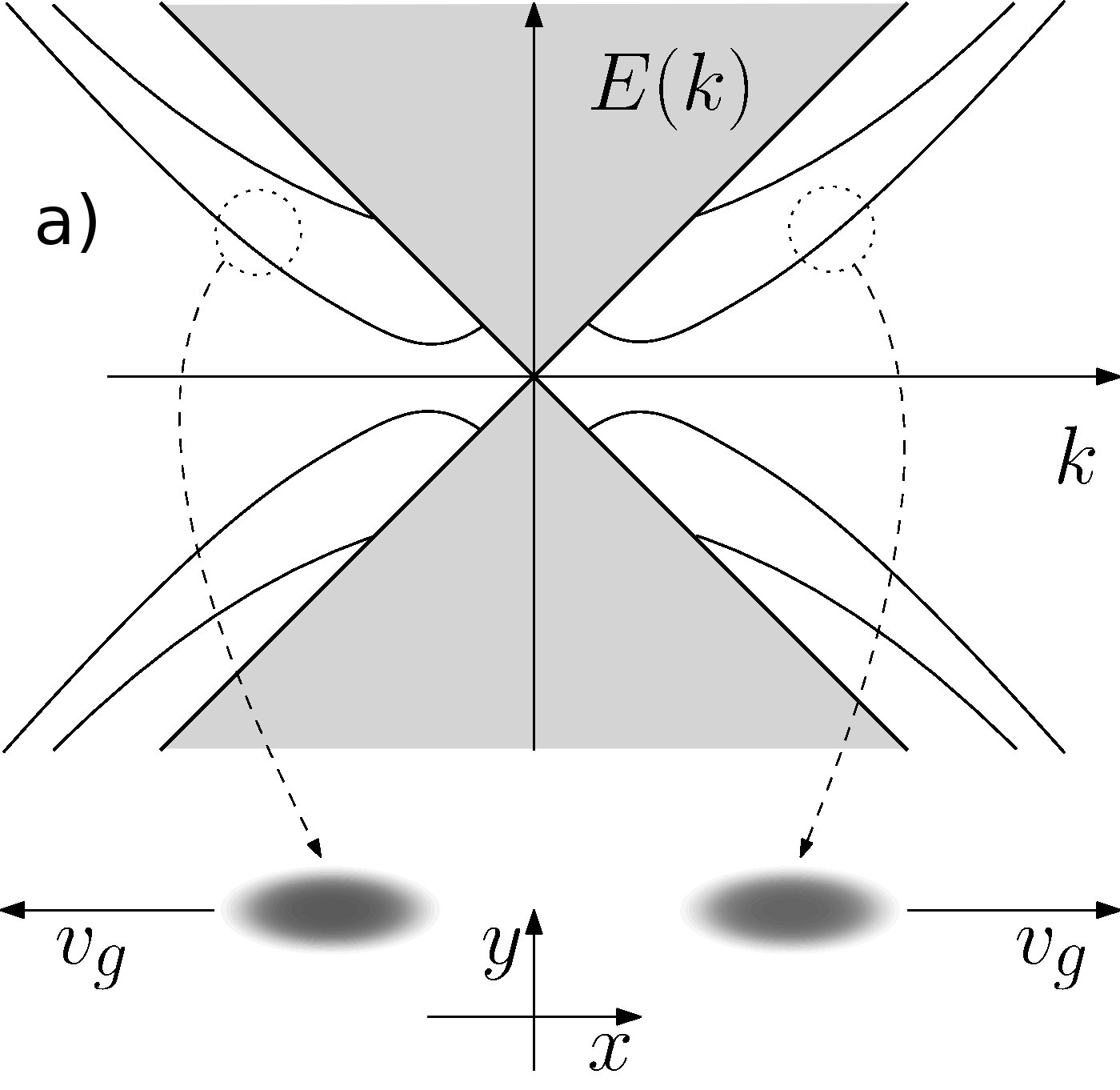}\hspace{8mm}
		\includegraphics[scale=.36]{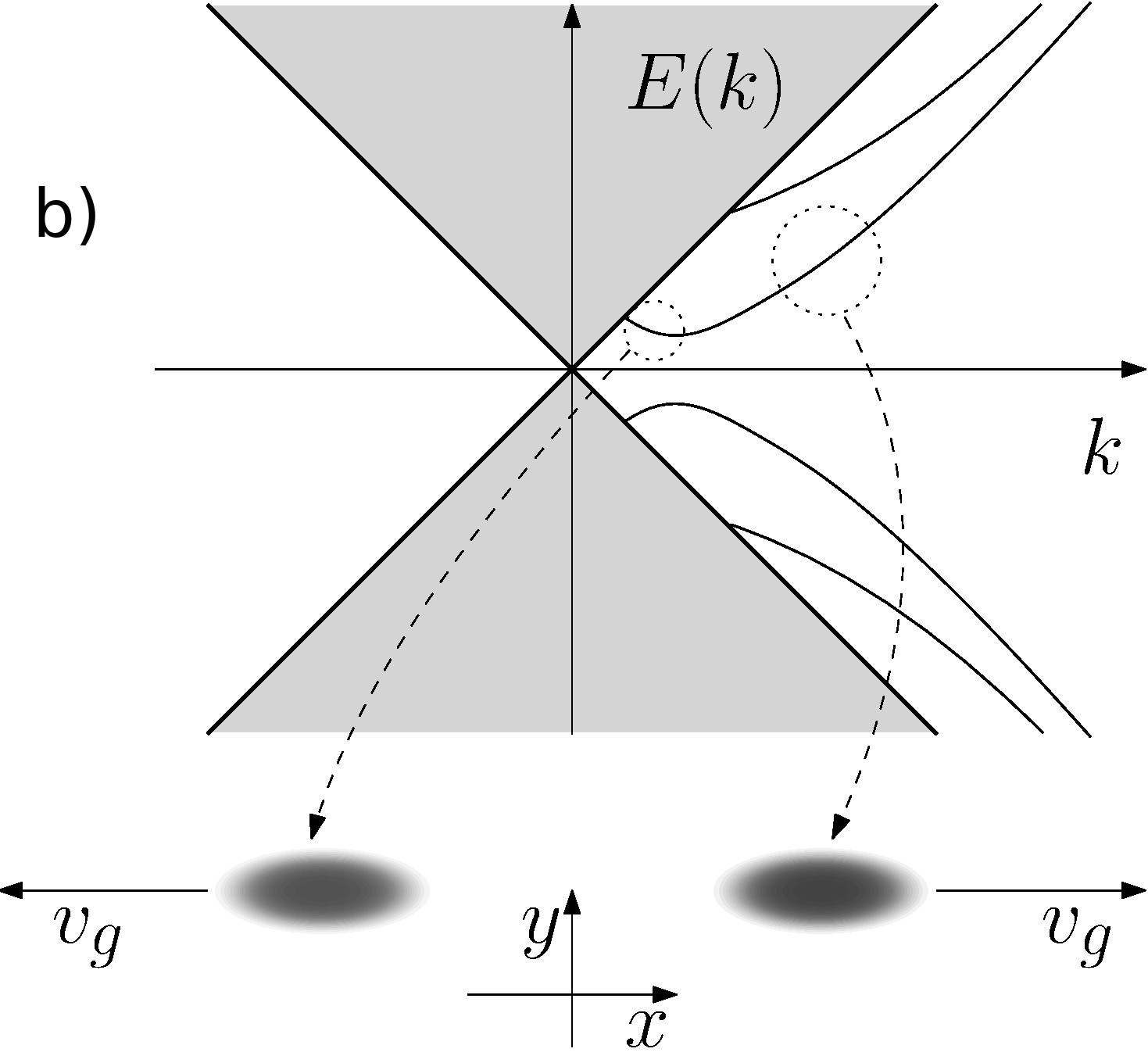}\hspace{8mm}			\includegraphics[scale=0.36]{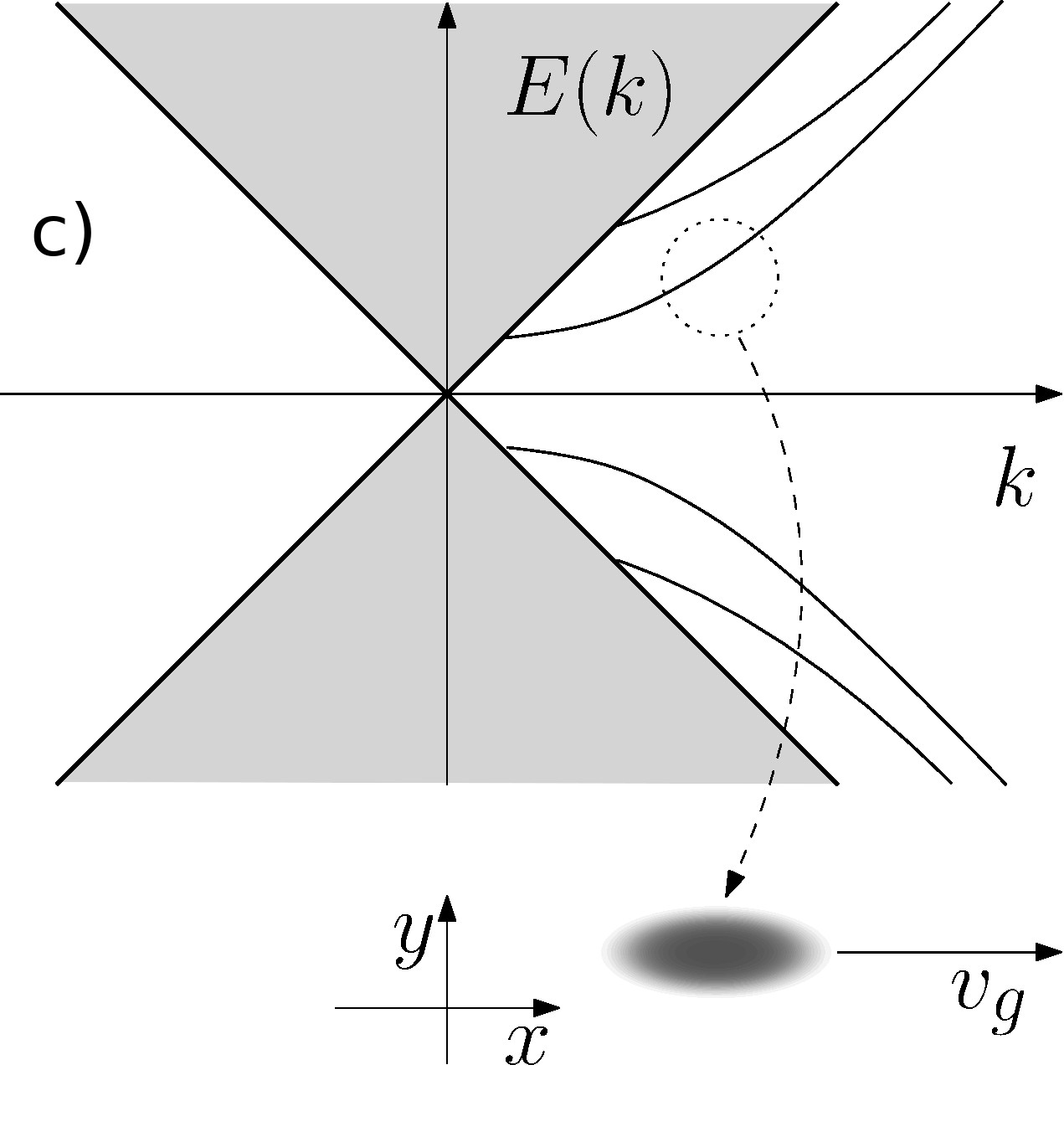}
		\end{center}
	\caption{Spectrum of the system described by $h(x,y)$. Discrete energy bands can be associated with the partially dispersionless wave packets (\ref{dispwp}) that are a) unidirectional  b) essentially unidirectional and c) unidirectional. Below the energy graph, there is illustration of the wave packets (\ref{dispwp}) composed from the states corresponding to the energies lying inside the dotted circles. The average group velocity of the wave packets is $v_g$ defined in (\ref{vg}). }\label{bands}
	\end{figure}
\end{center}

The speed of the wave packets can be approximated by the averaged group velocity 
\begin{equation}v_g=v_0\frac{\int_{I_n}\partial_k E_n(k)dk}{|\mathcal{I}_n|}.\label{vg}
\end{equation}
The spectrum of $H(x,y)$ is symmetric with respect to $0$.  The transport of the wave packets (\ref{dispwp}) is \textit{bidirectional} when $\partial_kE_n(k)$ of the positive energy bands\footnote{The negative energy bands has opposite sign of derivative so that the corresponding wave packets move in opposite direction. However, as they are composed of holes, they contribute to the same direction of electric current.} can be both positive and negative for $k\in\mathcal{I}_n$, see Fig.~\ref{bands}a). When $\partial_kE_n(k)$ of the positive energy bands is positive (negative) for all $k\in\mathcal{I}_n$, the corresponding wave packets defined in (\ref{dispwp}) can move in one direction only, they are \textit{unidirectional}, see Fig. \ref{bands}c). When the derivative of the positive energy bands has negative (positive) sign on a finite interval of $k$ and positive (negative, respectively) sign for all other $k\in\mathcal{I}_n$, we say that the transport is \textit{essentially unidirectional}, see Fig. \ref{bands}b). 

It is worth noticing that the construction of the partially dispersionless wave packets is not limited to our model but can be applied to broad class of systems with translational symmetry, see \cite{VJMT} for more details. Examples can be found in the literature where explicit models were solved numerically. See e.g. \cite{zhai}, \cite{wu} and \cite{villegas} for the discrete energy bands corresponding to the unidirectional, essentially unidirectional and bidirectional wave packets, respectively.   

Explicit solutions of (\ref{stacstate}) are needed for construction of the wave packets (\ref{dispwp}). There are well known exactly models where the stationary equation  (\ref{stacstate}) is exactly solvable, let us mention the model with $F(s)=\tanh s$. However, for reconstruction of the initial system described by $H(x,y)$, we would need the explicit form of $\mathbf{u}_{22}(y)$ (or $\mathbf{v}_{22}(y)$). It would be rather difficult to extract it from $F(s(y))=\tanh\left(\int_0^y \frac{v_0}{\mathbf{v}_{22}(q)}dq\right)=-\frac{\beta}{2a}\mathbf{u}_{22}(y)$.
Nevertheless, there is one scenario where we can get a partial solution immediately. When 
\begin{equation}
\lim_{s\rightarrow\pm\infty}F(s)=F_{\pm}, \quad |F_+|\neq |F_-|,
\end{equation}
i.e. the strain along $y$ axis is asymptotically constant.
Then we can obtain the following zero modes for any fixed $k$
\begin{equation}
\psi_+=e^{ikr}\left(e^{\int(k+F(s))ds},0\right)^t,\quad \psi_-=e^{ikr}\left(0,e^{-\int(k+F(s))ds}\right)^t,
\end{equation}
\begin{equation}
\tilde{H}(r,s)\psi_{\pm}(r,s)=0,
\end{equation}
where one of them is vanishing exponentially for $|x|\rightarrow\infty$ provided that $(F_-+k)(F_++k)<0$. This way, we get the zero energy $E_0(k)=0$ of $\tilde{H}(r,s)$. The wave packets (\ref{dispwp}) associated with this energy band do not move as its average group velocity would be identically zero.

\subsubsection*{Criteria for existence of the discrete energy bands}

Instead of looking for exactly solvable configurations of (\ref{eq2}), we focus on the qualitative spectral analysis of the system described by (\ref{h(r,s)}). It provides us with useful information on the energy bands $E_n(k)$ without the need to solve the stationary equation. We will find the sufficient conditions for the strain such that it gives rise to waveguides supporting unidirectional or bidirectional transport. 

Let us suppose that the second component of the strain tensor $\mathbf{u}_{22}(y)$ is asymptotically constant, $\lim_{|y|\rightarrow \infty}\mathbf{u}_{22}(y)=0.$ It implies that
\begin{equation}
\lim_{|y|\rightarrow \infty}\mathbf{v}_{22}(y)=v_0,\quad \lim_{|s|\rightarrow \infty}F(s)=0.
\end{equation}
We can utilize directly the results  presented in \cite{VJDK} and \cite{MFVJMT}. They are based on the fact that the square of $\tilde{H}_k(s)$ is Pauli-type diagonal Hamiltonian\footnote{In \cite{VJDK} and \cite{MFVJMT}, the considered Hamiltonian has permuted Pauli matrices when compared to (\ref{h(r,s)}), i.e. it would be in our notation $H(k,s)=-i\sigma_1\partial_s+(k+F(s))\sigma_2$. However, our application of the results of those works is insensitive to this change.}, $\tilde{H}_k^2(s)=-\partial_s^2+(k+F(s))^2+\sigma_3 F'(s)$, with the spectrum bounded from below. It is possible to use the variational principle to find criteria for existence of its discrete energies. Existence of discrete energies of $\tilde{H}_k^2(s)$ then implies existence of discrete energies in the spectrum of $\tilde{H}_k(s)$ and also of the associated discrete energy bands in the spectrum of $H(x,y)$. The corresponding eigenstates can be used in the construction of the dispersionless wave packets (\ref{stacstate})-(\ref{dispwp}).

Let us summarize some of the relevant criteria below:\\

\noindent
\textit{
	Let us suppose that $F(s)$ is continuous together with its first derivative and  $\int_{\mathbb{R}}|F(s)(F(s)+k)|ds<\infty$ or there exists $s_0$ such that $F(s)(F(s)+k)<0$ for all $s<s_0$. Then we can make the following conclusions \cite{MFVJMT}:
	\begin{enumerate}
		\item [1)] if $F(s)\geq 0$ (or $F(s)\leq0$) for all $s\in\mathbb{R}$, then there are no discrete energies in the spectrum of $\tilde{H}_k(s)$ for all $k\geq 0$ (or for all $k\leq0$, respectively).
		\item [2a)] if $\int^{s_0}_{-\infty}F(s)ds<0$ (or if $\int^{s_0}_{-\infty}F(s)ds>0$), then there exists $K>0$ such that for all $k>K$ (or for all $k<-K$, respectively) there are discrete energies in the spectrum of $\tilde{H}_{k}(s)$. 
		\item [2b)]if $\int_{s_0}^{\infty}F(s)ds<0$ (or if $\int_{s_0}^{\infty}F(s)ds>0$), then there exists $K>0$ such that for all $k>K$ (or for all $k<-K$, respectively) there are discrete energies in the spectrum of $\tilde{H}_{k}(s)$. 
	\end{enumerate}
	\begin{enumerate}
		\item [3a)] If $F(s)\geq 0$ and $F(s)\neq 0$ then $\tilde{H}_k(s)$ has discrete energies for all $k<-\frac{1}{2}\max_{s\in\mathbb{R}} F(s)$
		\item [3b)]  If $F(s)\leq 0$ and $F(s)\neq 0$ then $\tilde{H}_k(s)$ has discrete energies for all $k>\frac{1}{2}|\min_{s\in\mathbb{R}} F(s)|$
		\end{enumerate}
	Suppose that $F(s)$ is integrable.
		\begin{enumerate}
		\item [3c)]  If $\int_\mathbb{R}F(s)ds>0$ then there are discrete energy values in the spectrum of $\tilde{H}_k(s)$ for all $k<-\frac{\int_{\mathbb{R}}F(s)^2ds}{2\int_{\mathbb{R}}F(s)ds}$
		\item [3d)] If $\int_\mathbb{R}F(s)ds<0$ then there are discrete energy values in the spectrum of $\tilde{H}_k(s)$ for all $k>\frac{\int_{\mathbb{R}}F(s)^2ds}{2\left|\int_{\mathbb{R}}F(s)ds\right|}$
	\end{enumerate}  
}
A remark is in order. The criteria $3a)-3d)$ represent sufficient conditions for existence of the discrete energies of $\tilde{H}_k(s)$. It is possible that the discrete energy levels exist also outside of the specified interval for $k$. However, we do not have any tool how to guarantee their existence in that case. When $F(s)\geq 0$ ($F(s)\leq 0$), then $3a)$ and $3c)$ (or $3b)$ and $3d)$) provide two different values for the threshold values of $k$. It is not possible to decide which one provides better estimate without evaluating them for an explicit $F(s)$. 

\subsubsection*{Waveguides for essentially unidirectional wave packets}
We can use these criteria to show that the deformation with 
\begin{equation} \mathbf{u}_{22}\geq0\label{u22}\end{equation}
induces a wave guide for the essentially unidirectional wave packets (\ref{dispwp}). Indeed,
as we have $F(s(y))=-\frac{\beta}{2a}\mathbf{u}_{22}(y)$, there holds $F(s)\leq 0$. Then it follows from $1)$ that there are no discrete energies in the spectrum of $\tilde{H}_k(s)$ for any $k\leq0$.
We also know from $3b)$ and $3d)$ that there is a threshold $$k_0=\min\left\{\frac{\int_{\mathbb{R}}F(s)^2ds}{2\left|\int_{\mathbb{R}}F(s)ds\right|},\frac{1}{2}|\min_{s\in\mathbb{R}} F(s)|\right\}$$
such that the effective Hamiltonian $\tilde{H}_k(s)$ has discrete energies for all $k>k_0$. The value of $k_0$ can be expressed in terms of $\mathbf{u}_{22}(y)$ and the vertical  displacement $h=h(y)$ with the use of (\ref{rs}) in the following manner 
\begin{equation}
k_0=\min\left\{
\left|\frac{\beta}{4a}\right|\frac{\int_{\mathbb{R}}\frac{\mathbf{u}_{22}(y)^2}{1+(1-\beta) \mathbf{u}_{22}(y)-\frac{1}{2}(\partial_yh)^2}dy}{\left|\int_{\mathbb{R}}\frac{\mathbf{u}_{22}(y)}{1+(1-\beta) \mathbf{u}_{22}(y)-\frac{1}{2}(\partial_yh)^2}dy\right|},\frac{\beta}{4a}\max_{y\in\mathbb{R}}\mathbf{u}_{22}(y))|\right\},
\end{equation}
 
\vspace{1mm}

\noindent
\textit{We can conclude that whenever there holds $\mathbf{u}_{22}\geq0$ (but not identically $\mathbf{u}_{22}=0$), the strain induces a waveguide for essentially unidirectional transport of the dispersionless wave packets (\ref{dispwp}).} 

\vspace{1mm}
An inhomogeneous unidirectional strain  is a good example of the deformation that gives rise to (\ref{u22}).  The associated deformation vector is
\begin{equation}u=(-\mu\epsilon f(x),\epsilon f(y)),\quad f'(y)\geq0,\end{equation} 
where $\epsilon$ is the strain and $\nu$ is the Poisson ratio\footnote{The homogeneous unidirectional strain corresponds to $f(x)=x$.} gives rise to the following strain tensor and Fermi velocity  
\begin{equation}\label{bidirectionalstrain2}
 \mathbf{u}=\left(\begin{array}{cc}-\mu f'(x)&0\\0&\epsilon  f'(y)\end{array}\right),\quad \mathbf{v}=v_0\left(\begin{array}{cc}1-\mu(1-\beta)\epsilon f'(x)&0\\0&1+(1-\beta)\epsilon f'(y)\end{array}\right).
\end{equation}

\subsubsection*{Waveguides for bidirectional wave packets}
The waveguides formed by positive (negative) $\mathbf{u}_{22}(y)$ for all $y\in\mathbb{R}$ host essentially unidirectional wave packets. If we want to create waveguides that would host bidirectional transport, $\mathbf{u}_{22}$ has to acquire both positive and negative values. Revising the criteria $2a)$ and $2b)$, we can see that one way to create a waveguide for bidirectional wave packets is to have $\mathbf{u}_{22}\leq0$ of $y\in(-\infty,y_0)$ and $\mathbf{u}_{22}\geq0$ for $y\in(y_1,\infty)$.  The negative values of $\mathbf{u}_{22}$ mean that the atoms from the lattice have to get closer together. It is worth mentioning that free standing graphene is stable for small values of compressive strain  only. When the compression exceeds the threshold value of $\sim 0.1\%$, the strain gets compensated by creation of folds  \cite{si}. Hence, the experimental formation of waveguides for bidirectional wave packets by compressive strain could be a rather complicated task.

\subsubsection*{Waveguides for valleytronics}
Let us suppose that strained graphene with Fermi velocity (\ref{bidirectionalstrain}) is in presence of an external magnetic field perpendicular to the crystal,  $B=(0,0,\partial_y A_x(y))$. Contrary to the pseudo-magnetic gauge field induced by the mechanical deformations, the magnetic field breaks the time-reversal symmetry and comes with opposite sign when we consider Dirac fermions in the vicinity of the second Dirac point $\mathbf{K'}\equiv\mathbf{-K}$. The Dirac Hamiltonians at the Dirac points $\mathbf{K}$ and $-\mathbf{K}$ can be written in the following form
\begin{equation}
H_{\pm \mathbf{K}}=\mp i\sigma_1\sqrt{\mathbf{v}_{11}(x)}\partial_x\sqrt{\mathbf{v}_{11}(x)}-i\sigma_2\sqrt{\mathbf{v}_{22}(x)}\partial_x\sqrt{\mathbf{v}_{22}(x)}\pm\sigma_1\frac{v_0\beta}{2a}(\mathbf{u}_{11}(x)-\mathbf{u}_{22}(y))+ \frac{v_0\beta}{2a}\sigma_1A_x(y).
\end{equation}
Let us set the strain and the magnetic field such that 
\begin{equation}
A_x(y)=\mathbf{u}_{22}(y).
\end{equation}
Then we follow the steps (\ref{gauge})-(\ref{tildeh(k,s)}) for both $H_{\pm\mathbf{K}}$ and get the effective one dimensional operators 
\begin{eqnarray}
\tilde{H}^{\mathbf{K}}_k(s)&=&-i\sigma_2\partial_s+\sigma_1 k,\\
\tilde{H}^{\mathbf{K'}}_k(s)&=&-i\sigma_2\partial_s+\sigma_1\left(-k+\frac{\beta}{a}\mathbf{u}_{22}(y(s))\right).
\end{eqnarray}
Therefore, the Dirac fermions at the vicinity of the Dirac point $\mathbf{K}$ are effectively governed by free-particle Hamiltonian which has no discrete energies in its spectrum. On the other hand, the Dirac fermions at the Dirac point $\mathbf{K}'$ are subject to the vector potential $\frac{v_0\beta}{a}\mathbf{u}_{22}(y)$ that can induce discrete energies. In particular, when $\mathbf{u}_{22}\geq 0$, the strain forms waveguide for essentially unidirectional wave packets in the $\mathbf{K}'$-valley whereas the combination of the strain and the magnetic field does not confine Dirac fermions in the $\mathbf{K}$-valley.  The combination of the external magnetic field with the strain in order to control propagation of the electrons in $\mathbf{K}$ and $\mathbf{K}'$ valleys appeared e.g. in construction of valley filters \cite{wupeeters}, \cite{yesilyurt},  \cite{chaves}.

\section*{Discussion}
We showed that the Fermi velocity barrier induced by the strain (\ref{v1}) gives rise to deflection of the incoming wave packets and possible focusing of the wave packets, see Fig.~\ref{bla}. The shifted trajectory as well as the deflection angle can be found explicitly (\ref{n}), (\ref{omega}). This effect is similar to the Goos-H\"anchen-like effect for transmitted electrons discussed e.g. in \cite{wupeeters}, \cite{cao}, \cite{yesilyurt}, \cite{zhai2}, \cite{zhang}, yet there is no reflection on the barrier in our case. The reason is that our system can be mapped to the free particle model where the wave packets do not suffer from any scattering. It is worth mentioning that similar situation was discussed in \cite{VJMP} where unitary mapping to free particle model was used to explain the absence of backscattering of Dirac fermions on the impurities in carbon nanotubes or on the electrostatic barriers. It resembles the P\"oschl-Teller reflectionless system known in non-relativistic quantum mechanics that can be related to the free particle system by Darboux transformation \cite{PoschTellerDarboux}.

In the section three, we discussed how deformations represented by the strain tensor (\ref{bidirectionalstrain}) can induce waveguides for partially dispersionless wave packets (\ref{dispwp}). 
We used the results of qualitative spectral analysis. The wave packets (\ref{dispwp}) associated with the discrete energy bands can have major influence on the conduction of the waveguide as they do not disperse rapidly outwards the wave guide during time evolution. We found that any nonvanishing deformation (\ref{bidirectionalstrain}) with $\mathbf{u}_{22}\geq0$ gives rise to the wave guide for the essentially unidirectional wave packets. Our results are complementary to the existing literature where explicit models were considered. Guided modes in the waveguides induced by inhomogeneous Fermi velocity were discussed e.g. in \cite{wang}, \cite{yuan}, \cite{portnoi}, \cite{roy}  with Fermi velocity fixed as $\mathbf{v}=v(x)\eta$. In \cite{zhai}, \cite{pellegrino}, the Fermi velocity was associated with the applied strain. These models differ from our one by presence of external fields (typically electric potential) or by the Fermi velocity that appears in the Hamiltonian without the associated pseudo-magnetic vector potential (\ref{A}). 

Finally, let us notice that despite we supposed the Dirac fermion to move in graphene throughout the work, there is an expanding family of Dirac materials where dynamics of low-energy particles is described by the same equations \cite{organic}, \cite{molecular}, \cite{wehling}, \cite{polini}. This broadens the applicability of the obtained results to a wider class of physical systems.

\appendix
\section{Properties of the dispersionless wave packets}
The norm of the wave packet \eqref{dispwp} is given by the coefficient function $\rho$.
\begin{eqnarray}
&&\int_{\mathbb{R}^2}\left|\Psi(x,y)\right|^2dxdy=\int_{\mathbb{R}^2}\left|\frac{1}{\sqrt{2\pi}}\frac{v_0 ~e^{-i\frac{v_0\beta}{2a}\int \mathbf{u}_{11}/\mathbf{v}_{11}dx}}{\sqrt{\mathbf{v}_{11}\mathbf{v}_{22}}}\int_{\mathcal{I}}\rho(k)e^{ikr(x)}\psi_k(s(y))dk\right|^2dxdy\nonumber\\&&=\int_{\mathbb{R}^2}\left|\frac{1}{\sqrt{2\pi}}\int_{\mathcal{I}}\frac{v_0~e^{ikr(x)}\rho(k)}{\sqrt{\mathbf{v}_{11}(x)\mathbf{v}_{22}(y)}}\psi_k(s(y))dk\right|^2dxdy=\int_{\mathbb{R}^2}\left|\frac{1}{\sqrt{2\pi}}\int_{\mathcal{I}}e^{ikr}\rho(k)\psi_k(s)dk\right|^2drds\nonumber\\
&&=\int_{\mathbb{R}^2}\left|\rho(k)\psi_k(s)\right|^2dkds=\int_{\mathbb{R}}\left|\rho(k)\right|^2dk.
\end{eqnarray} 
On the last line, we used the fact that the Fourier transform is a unitary mapping and $\psi_k(s)$ is a normalized bound state of $\tilde{H}_k(s)$.

The wave packets (\ref{dispwp}) do not disperse in $y$ direction.  In what follows, we do not write the label of the energy band explicitly, $E(k)\equiv E_n(k)$, $\mathcal{I}\equiv \mathcal{I}_n$, $\rho(k)\equiv\rho_n(k)$, $\psi_{k}(s)\equiv\psi_{n,k}(s)$. The probability of finding the particle in the interval $y\in(a,b)$ at time $t$ can be calculated as
\begin{eqnarray}
&&\int^{b}_a\int_{\mathbb{R}}\left|e^{-itH(x,y)}\Psi(x,y)\right|^2dxdy=\int^{{b}}_{{a}}\int_{\mathbb{R}}\left|G^{-1}(x,y)e^{-it\tilde{H}(x,y)}G(x,y)\frac{1}{\sqrt{2\pi}}\int_{\mathcal{I}}\rho(k)G^{-1}(x,y)e^{ikr(x)}\psi_k(s(y))dk\right|^2dxdy\nonumber\\
&&=\int^{\tilde{b}}_{\tilde{a}}\int_{\mathbb{R}}\left|e^{-it\tilde{H}(r,s)}\frac{1}{\sqrt{2\pi}}\int_{\mathcal{I}}\rho(k)e^{ikr}\psi_k(s)dk\right|^2drds
=\int^{\tilde{b}}_{\tilde{a}}\int_{\mathbb{R}}\left|\frac{1}{\sqrt{2\pi}}\int_{\mathcal{I}}e^{-itE(k)+ikr}\rho(k)\psi_k(s)dk\right|^2drds
\nonumber\\
&&=\int^{\tilde{b}}_{\tilde{a}}\int_{\mathcal{I}}\left|e^{-itE(k)}\rho(k)\psi_k(s)\right|^2dkds
=\int^{\tilde{b}}_{\tilde{a}}\int_{\mathcal{I}}\left|\rho(k)\psi_k(s)\right|^2dkds
=\int^{\tilde{b}}_{\tilde{a}}\int_{\mathbb{R}}\left|\frac{1}{\sqrt{2\pi}}\int_{\mathcal{I}}\rho(k)e^{ikr}\psi_k(s)dk\right|^2drds
\nonumber\\
&&=\int^{{b}}_{{a}}\int_{\mathbb{R}}\left|\frac{1}{\sqrt{2\pi}}\int_{\mathcal{I}}\rho(k)e^{ikr(x)}\frac{v_0 ~\psi_k(s(y))}{\sqrt{\mathbf{v}_{11}(x)\mathbf{v}_{22}(y)}}dk\right|^2dxdy=\int_{a}^b\int_{\mathbb{R}}\left|\Psi(x,y)\right|^2dxdy,
\end{eqnarray}
where we used unitarity of the Fourier transform on the third line. We can see that it does not change in time. Since $a$ and $b$ arbitrary, we can conclude that the probability of finding the particle in a fixed interval of the $y$ axis does not change in time. 

\section*{Acknowledgements}
V.J. was supported by GA\v CR grant no.19-07117S.

\end{document}